\documentclass[aps,prd,reprint,groupedaddress,showpacs,nofootinbib,amsmath,onecolumn,eqsecnum]{revtex4-2}

\usepackage{amsmath}
\usepackage{amssymb}
\usepackage{bbold}
\usepackage{ulem}
\usepackage{graphicx}
\usepackage{verbatim}
\usepackage{hyperref}
\usepackage{indentfirst}
\usepackage{mathtools}
\usepackage{xcolor}
\usepackage{color}
\usepackage{adjustbox}
\usepackage{placeins}
\usepackage{lipsum}
\usepackage{csquotes}

\usepackage{calligra}

\usepackage[font=small,labelfont=bf,labelsep=space]{caption}
\newcommand{\md}{\mathrm{d}}

\begin{document}

%Sorry to tweak the format. I get more motivated if it looks like a paper from the start.

\title{Disformal symmetry in the Universe: mimetic gravity and beyond}

\author{\textsc{Guillem Dom\`enech$^{a}$}}
    \email{{guillem.domenech}@{itp.uni-hannover.de}}

\author{\textsc{Alexander Ganz$^{b}$}}
    \email{{alexander.ganz}@{uj.edu.pl}}

\affiliation{$^a$ Institute for Theoretical Physics, Leibniz University Hannover, Appelstraße 2, 30167 Hannover, Germany.}

\affiliation{$^b$ Faculty of Physics, Astronomy and Applied Computer Science, Jagiellonian University, 30-348 Krakow, Poland}

\begin{abstract}
Symmetries play an important role in fundamental physics. In gravity and field theories, particular attention has been paid to Weyl (or conformal) symmetry. However, once the theory contains a scalar field, conformal transformations of the metric can be considered a subclass of a more general type of transformation, so-called disformal transformation. Here, we investigate the implications of pure disformal symmetry in the Universe. We derive the form of general disformal invariant tensors from which we build the most general disformal invariant action. We argue that, in cosmology, disformal symmetry amounts to require that the lapse function is fully replaced by a (time-like) scalar field at the level of the action. We then show that disformal symmetry is in general an exactly equivalent formulation of general mimetic gravity. Lastly, we go beyond mimetic gravity and find that a particular class of invariance leads to seemingly Ostrogradski-like (with higher derivatives) Lagrangians, which are nevertheless absent of Ostrogradski ghosts in a cosmological background, despite having an additional degree of freedom. We also propose an application of our formalism to find new invertible disformal transformations, where the coefficient involves higher derivatives and curvature, further expanding the theory space of scalar-tensor theories.
\end{abstract}
\maketitle

\section{Introduction}

Symmetries, and breaking thereof, act as a guiding principle from which one can understand and build a fundamental theory, as in, e.g., the standard model of particle physics. Symmetries also play an important role in cosmology. In the very early universe (and the recent one), the Universe experienced an almost de Sitter expansion which led to an almost scale invariant spectrum of primordial fluctuations \cite{Planck:2018vyg,Planck:2018jri}. One could argue that perhaps this is an indication that scale symmetry is a property of quantum gravity \cite{Rubio:2017gty,Wetterich:2019qzx,Wetterich:2020cxq}, both at high and low energies. It has also been noticed that the boundary of de Sitter also enjoys a conformal symmetry which is used in the cosmological bootstrap program \cite{Arkani-Hamed:2018kmz}. Conformal symmetry is also key in the AdS/CFT correspondence. However, once the theory contains a scalar field, as is usually the case in cosmology, one can consider more general transformations of the metric. The so-called disformal transformations were introduced by Bekenstein \cite{Bekenstein:1992pj} and involve derivatives of a scalar field. Conformal transformations are a subclass of disformal transformations. 

Departing from the de Sitter symmetry, we may use that the Universe has a preferred direction of time to argue that time diffeomorphism is, in fact, ``broken'' in cosmology \cite{Cheung:2007st}. One can then write down those theories which are compatible with the remaining spatial diffeomorphism such as the Effective Field Theory (EFT) of inflation \cite{Cheung:2007st,Weinberg:2008hq} (see also \cite{Arkani-Hamed:2003pdi,Arkani-Hamed:2003juy} for the first works on the context of Ghost inflation), dark energy \cite{Gubitosi:2012hu,Kase:2014cwa,Tsujikawa:2014mba,Langlois:2017mxy,Frusciante:2019xia} and in general spatially covariant gravity \cite{Gao:2014fra,Gao:2014soa}. Disformal transformations have played a key role in building and understanding general scalar-tensor theories of gravity (such as Horndenski \cite{Horndeski:1974wa}, beyond Horndeski \cite{Gleyzes:2014dya}, DHOST \cite{Langlois:2015cwa}, U-DHOST \cite{DeFelice:2018ewo} and beyond--see \cite{Langlois:2018dxi,Kobayashi:2019hrl} for reviews) as well as their related aforementioned EFTs (see, e.g., refs.~\cite{Zumalacarregui:2013pma,Domenech:2015tca,Fujita:2015ymn,Crisostomi:2016tcp,BenAchour:2016cay,BenAchour:2016fzp,Takahashi:2017zgr,BenAchour:2020wiw,Babichev:2021bim} and references therein).  However, to the best of our knowledge, the role and meaning of disformal symmetry in the Universe has not yet been explored. Here we will investigate \textit{pure} disformal symmetry and disregard global conformal factors. In this way, we shall understand the implications of disformal symmetry without mixing of conformal symmetry.

Pure disformal symmetry in cosmology, in its most naive level, implies that a theory is invariant under a rescaling of the time coordinate or, alternatively, the lapse function \cite{Domenech:2015hka}. In the EFT language (or in the uniform scalar field slicing), disformal symmetry implies a ``lapse-less'' action. But, this point of view is not very informative and a ``lapse-less'' action sounds more worrisome, in terms of the standard Hamiltonian analysis, than it actually is. A more insightful perspective is that pure disformal symmetry demands that ``time'' is fully represented by a scalar field at the level of the action (not the metric). Thus, rescaling of time (or the lapse) is not important because the fundamental field describing time is such a scalar.

To convince the reader about this point, consider the (3+1) decomposition in which the coordinate time $t$ is always accompanied by the lapse function $N$, namely as $Ndt$. If the lapse $N$, for some reason, were related to the time derivative of a fundamental scalar field, say through $N=\dot\phi$, we would instead have $\dot\phi dt=d\phi$. This is very roughly a consequence of pure disformal symmetry but at the level of the action. In passing, we recall that a constraint imposing $N=\dot\phi$ appears in mimetic gravity \cite{Chamseddine:2013kea}. As we shall show, a generic disformal symmetric theory is general mimetic gravity \cite{Arroja:2015wpa,Takahashi:2017pje,Langlois:2018jdg} in disguise. This is, of course, a simplified view and one can generalize the formalism to go beyond mimetic gravity.

Mimetic gravity was first introduced in \cite{Chamseddine:2013kea} (see also \cite{Chamseddine:2014vna,Chaichian:2014qba,Mirzagholi:2014ifa}) as a possible candidate to explain dark matter. Since then, it has built a long literature with applications to cosmology, dark energy, darkmatter, inflation and black holes (see \cite{Sebastiani:2016ras} for a review). Although mimetic gravity was first introduced via a degenerate conformal transformation, it may also be invoked directly with a Lagrange multiplier \cite{Golovnev:2013jxa,Barvinsky:2013mea} and via a degenerate disformal transformation \cite{Deruelle:2014zza,Arroja:2015wpa,Jirousek:2022jhh,Jirousek:2022rym}. For mimetic gravity within DHOST see \cite{Langlois:2018jdg}. Mimetic gravity also appears in noncommutative geometry \cite{Chamseddine:2014nxa}.  For recent works on mimetic gravity from the Hamiltonian and EFT points of view see \cite{Ganz:2018mqi,Ganz:2019vre}. The symmetries of the mimetic gravity action without Lagrange multiplier is discussed in \cite{Hammer:2015pcx} for conformal mimetic and in general in \cite{Jirousek:2022jhh,Golovnev:2022jts}. In particular, \cite{Jirousek:2022jhh} noticed that mimetic gravity via degenerate disformal transformations is invariant under generic disformal transformations of the auxiliary metric. However, this does not imply that all disformal invariant actions belong to mimetic gravity. For instance, not all conformal invariant actions correspond to mimetic gravity.

In this paper, we start from first principles and require disformal symmetry of the action. Our formulation does not depend on any degenerate metric transformation. We then build general disformal invariant tensors, generalizing the disformal invariant curvature tensor constructed in \cite{Domenech:2019syf}, from which we build a disformal symmetric action. We will later see that in general a disformal symmetric action leads to mimetic gravity. By relaxing the disformal symmetry we can go beyond mimetic gravity. These are mimetic-like theories, within spatially covariant gravity with a dynamical lapse \cite{Gao:2018znj,Lin:2020nro}, but contain a healthy extra degree of freedom. At the end, we propose  applications of disformal invariant tensors to higher derivative-dependent disformal transformations \cite{Babichev:2021bim,Takahashi:2021ttd,Naruko:2022vuh,Takahashi:2022mew,Ikeda:2023ntu,Takahashi:2023jro}, leading to new theories.

This paper is organized as follows. In \S\ref{sec:generic} we present a general disformal invariant action in both the covariant form and in the unitary gauge. We then prove that these class of actions correspond to mimetic gravity. In \S\ref{sec:restricted} we allow for more general actions by considering a relaxed disformal symmetry. In \S\ref{sec:cosmo} we study the cosmology of the system including linear perturbations. In \S\ref{sec:highercurvature} we show that, by adding higher curvature terms allowed by the disformal symmetry, the mimetic-like degree of freedom can be rendered healthy. We conclude our work and discuss future directions in \S\ref{sec:conclusions}. Details of the formulas can be found in the appendices.

\newpage

\section{Invariance under generic disformal transformation \label{sec:generic}}

Let us consider that our gravity theory contains a scalar field $\phi$. A pure disformal transformation is defined by \cite{Bekenstein:1992pj}
\begin{align}\label{eq:disf}
g_{\mu\nu} \to g_{\mu\nu}+ D(x^\mu)\partial_\mu\phi\partial_\nu\phi\,,
\end{align}
where we have disregarded a global conformal factor, which would eventually connect with conformal invariance, and we allowed a general form of the disformal factor $D$. It is often assumed that $D=D(\phi,X)$ but we will require \textit{generic} invariance.

From now on, we shall focus on a cosmological situation, where the scalar field is time-like and offers a preferred foliation of the space-time, i.e. a slicing of uniform $\phi$. The same follows for a space-like scalar field with  care of the proper signs. The orthogonal vector to such foliation is given by
\begin{align}\label{eq:constant-phi}
n_\mu = \frac{\partial_\mu\phi}{\sqrt{X}}\quad{\rm with}\quad
X=-g^{\mu\nu}\partial_\mu\phi\partial_\nu\phi\,,
\end{align}
and the metric in such $3+1$ decomposition reads
\begin{align}\label{eq:ginh}
g_{\mu\nu}=h_{\mu\nu}-n_\mu n_\nu\,,
\end{align}
where $h_{\mu\nu}$ is the intrinsic metric of the spatial 3 dimensional hypersurface. In the uniform-$\phi$ slicing, the transformation \eqref{eq:disf} is readily understood as a rescaling of the orthogonal vector, namely
\begin{align}\label{eq:n_mu}
n_\mu \to \sqrt{1-DX}n_\mu\,.
\end{align}

Before going into the details, it is instructive to consider $D=D(\phi,X)$ and intuitively anticipate what will follow. For $D(\phi,X)$, invariance under the disformal transformation \eqref{eq:disf} can be recast as invariance under rescaling of the orthogonal vector \eqref{eq:n_mu} up to field redefinitions. One way to do this is to require invariance of the orthogonal vector itself, which implies $1-DX=F(\phi)$ in \eqref{eq:n_mu}. The function $F(\phi)$ can be absorbed into a field redefinition of $\phi$. We then have two possibilities: either there is a constraint imposing $X=f(\phi)$ (e.g. see \cite{Lim:2010yk}) or $D= d(\phi)/X$. The first case corresponds to general mimetic gravity \cite{Chamseddine:2013kea,Langlois:2018jdg} which we discuss in \S\ref{subsec:mimetic}. The second allows for a healthy Ostrogradski ``ghost'' \cite{Ganz:2020skf} or, more precisely, a healthy additional degree of freedom. We postpone the second case until \S\ref{sec:restricted}. We will prove these arguments explicitly at the action level.

We now proceed to construct general actions invariant under generic disformal transformations. We will work in the covariant formulation as well as in the unitary gauge. The notation ``unitary gauge'', as often used in the EFT language, essentially corresponds to the uniform-$\phi$ slicing, i.e. $\delta\phi=0$ in cosmological perturbation theory, plus the additional fixing of $\phi=t$ by using time reparametrisation invariance or, in other words, by fixing the lapse function. However, in disformal symmetric theories the lapse is absent in the action and, to be consistent, we will keep track of $\dot\phi$ in most of our discussions. We will later see that the additional fixing of $\phi=t$ is compatible with disformal symmetry. As we shall see, although the calculations are simpler in the unitary gauge, the covariant formulation provides crucial insight in the case of mimetic gravity.

\subsection{Covariant formulation\label{sec:covariant}}

Let us start by writing a general form of the action that has disformal symmetry. We will do so using basic building blocks: disformal invariant tensors. Let us respectively call ${\cal V}$, ${\cal B}_{\mu\nu}$ and ${\cal D_{\alpha\beta\mu\nu}}$ the disformal invariant volume element, the disformal invariant second derivative of the scalar field $\phi$ and the disformal invariant curvature tensor. In addition, we need a disformal invariant ``metric'' or projector, call it ${\cal P}^{\mu\nu}$. The action then reads
\begin{align}\label{eq:actioncovariant0}
S_{\rm gen.D}=\int d^4x {\cal V}\Big\{&c_0(\phi)+c_1(\phi){\cal B}+c_2(\phi){\cal P}^{\alpha\mu}{\cal P}^{\beta\nu}{\cal D}_{\alpha\beta\mu\nu}\nonumber\\&+c_3(\phi){\cal B}^2+c_4(\phi){\cal P}^{\alpha\mu}{\cal P}^{\beta\nu}{\cal B}_{\alpha\beta}{\cal B}_{\mu\nu}+\ldots\Big\}\,,
\end{align}
where $c_i$ are functions of the field only (which is automatically disformal invariant) and the $\ldots$ refer to higher order contractions. The disformal invariant tensors in terms of $\phi$ and $X$ are given by
\begin{align}
{\cal V}&=\sqrt{- g X}\,,\label{eq:volume}\\
{\cal P}^{\mu\nu}&= g^{\mu\nu}+\frac{1}{X}\nabla^\mu\phi\nabla^\nu\phi \,,\label{eq:P}\\
{\cal B}_{\mu\nu}&=\frac{1}{X}\left(\nabla_\mu\nabla_\nu\phi-\frac{1}{X}\nabla_{(\mu}X\nabla_{\nu)}\phi-\frac{1}{2X^2}\nabla_\alpha \phi\nabla^\alpha X\nabla_\mu\phi\nabla_\nu\phi \right)\,,\label{eq:B}\\
{\cal D}_{\alpha\beta\mu\nu}&= R_{\alpha\beta\mu\nu}-\frac{2}{X}\nabla_{\mu}\nabla_{[\alpha}\phi\nabla_{\beta]}\nabla_{\nu}\phi-\frac{2}{X^2}
\nabla_{[\alpha}\phi\nabla_{\beta]}\nabla_{[\mu}X\nabla_{\nu]}\phi
-\frac{4}{X^3}\nabla_{[\alpha}\phi\nabla_{\beta]}X\nabla_{[\mu}\phi\nabla_{\nu]}X\nonumber\\&-\frac{2\nabla_\mu\phi\nabla^\mu X}{X^3}
\nabla_{[\alpha}\phi\nabla_{\beta]}\nabla_{[\mu}\phi\nabla_{\nu]}\phi\,,\label{eq:D}
\end{align}
where (anti)-symmetrization is normalised. In the action \eqref{eq:actioncovariant0} we have defined
\begin{align}
    {\cal B}\equiv {\cal P}^{\mu\nu}{\cal B}_{\mu\nu}\,.
\end{align}
One can check that these objects are disformal invariant using the formulas provided in appendix \ref{sec:disfforumlas}. There is also the disformal invariant contravariant vector given by
\begin{align}
    {\cal N}^\mu=& \frac{1}{{X}}\nabla^\mu\phi\,.
\end{align}
However, contractions of ${\cal N}^\mu$ with ${\cal P}_{\mu\nu}$ \eqref{eq:P} and ${\cal B}_{\mu\nu}$ \eqref{eq:B} trivially vanish and the non-vanishing contraction with ${\cal D}_{\alpha\beta\mu\nu}$ \eqref{eq:D}, that is ${\cal P}^{\alpha\mu}{\cal N}^{\beta}{\cal N}^\nu{\cal D}_{\alpha\beta\mu\nu}$, can be written in terms of the contractions appearing in the action \eqref{eq:actioncovariant0}, after integration by parts. There is also the trivial disformal invariant contraction ${\cal N}^\mu \nabla_\mu\phi=-1$ which corresponds to the term $c_0$ in \eqref{eq:actioncovariant0}. In the same appendix we provide the explicit expression of the contractions appearing in \eqref{eq:actioncovariant0}. Note that in \eqref{eq:volume} we are implicitly assuming that $X>0$. One could also treat the case $X<0$ by appropriately choosing the signs. But, in this disformal invariant formulation, $X=0$ is ill-defined. This is first indication that the system will not cross $X=0$ and we will likely have a constraint imposing at least $X>0$. That being said, one could also build a disformal invariant action only in the exact case where $X=0$. We leave this case for future study.

For easier comparison with scalar-tensor theories of gravity, such as DHOST \cite{Langlois:2017mxy}, we also write down  \eqref{eq:actioncovariant0} in terms of $\phi$, $X$ and the Ricci scalar $R$. After some integration by parts in \eqref{eq:actioncovariant0}, we obtain
\begin{align}\label{eq:actioncovariant}
S_{\rm gen.D}=\int d^4x \sqrt{- g X}\left\{c_0+(c_1-2Xc_{2,\phi})D_1+c_2D_2+c_3D_3+c_4D_4+\ldots\right\}\,,
\end{align}
where we dropped the explicit $\phi$ dependence in $c_i$, the subscript ``$,\phi$'' refers to derivative w.r.t. $\phi$ and we defined
\begin{align}
D_1&=\frac{1}{X}\left(\Box\phi-\frac{1}{2X}\nabla_\mu X\nabla^\mu \phi\right)\,,\\
D_2&=R+\frac{1}{X}\left((\Box\phi)^2-\nabla_\mu\nabla_\nu\phi\nabla^\mu\nabla^\nu\phi\right)\,,\\
D_3&=D_1^2\,,\\
D_4&=\frac{1}{X^2}\left(\nabla_\mu\nabla_\nu\phi\nabla^\mu\nabla^\nu\phi+\frac{1}{2X}\nabla_{\mu}X\nabla^{\mu}X+\frac{1}{4X^2}\left(\nabla_\mu \phi\nabla^\mu X\right)^2\right)\,.
\end{align}
We find that the action \eqref{eq:actioncovariant} up to quadratic and cubic orders is a particular case of DHOST \cite{Langlois:2017mxy}. We write our theory explicitly in terms of DHOST coefficients in appendix \ref{sec:U-DHOST}. Note, however, that generic disformal symmetry allows for higher order operators beyond cubic order in the action, which do not belong to DHOST.

If we  further require that disformal symmetry is also preserved in any matter sector (scalar, vectors, fermions, etc.), all matter fields must minimally couple to an effective metric which has a well-defined inverse. The only possibility is given by
\begin{align}\label{eq:gmatter}
g^{\rm matter}_{\mu\nu}={\cal P}_{\mu\nu}-\nabla_\mu\phi\nabla_\nu\phi\quad{,\quad } g^{\mu\nu}_{\rm matter}={\cal P}^{\mu\nu}-{\cal N}^{\mu}{\cal N}^{\nu}\,,
\end{align}
which as we shall shortly see it is nothing but the degenerate metric leading to mimetic gravity \eqref{eq:mimeticg}. In addition to \eqref{eq:mimeticg}, matter fields can additionally couple to ${\cal P}^{\mu\nu}$ and ${\cal N}^\mu$ without spoiling the disformal symmetry. And, in particular, extra scalar fields, can couple to ${\cal P}^{\mu\nu}$ and ${\cal N}^\mu$ without explicit mention to \eqref{eq:gmatter}, if no higher derivatives are present in the action. We now proceed to show that one can reach the same construction from the unitary gauge.

\subsection{Unitary gauge}
It is also useful to consider the construction of the disformal invariant theories in the spatial covariant gravity using the unitary gauge. 
Using the standard ADM decomposition
\begin{align}
    \md s^2 = - N^2 \md t^2 + h_{ij} \left( N^i \md t + \md x^i \right) \left( N^j \md t + \md x^j\right)\,,
\end{align}
the disformal transformation corresponds to
\begin{align}
    h_{ij} \rightarrow h_{ij}, \qquad N^k \rightarrow N^k, \qquad N \rightarrow N\tilde D(x^\mu)\,,
\end{align}
where we defined $\tilde D^2(x^\mu)={1+D(x^\mu)\dot\phi^2/N^2}$.
Thus, disformal symmetry corresponds to invariance under generic rescaling of the lapse function. 
We can now start from spatial covariant gravity and construct the invariant action as
\begin{align}\label{eq:actionunitary}
     S_{\rm gen.D} = \int \md^3x\,\md t\, \sqrt{h} \dot \phi \left( c_0 -  \frac{c_1}{\dot \phi} E + c_2 R[h] + \frac{c_3}{\dot \phi^2} E_{ij} E^{ij} + \frac{c_4}{\dot \phi^2} E^2 + ... \right)\,,
\end{align}
where $E_{ij}$ is the rescaled extrinsic curvature 
\begin{align}
    E_{ij} = N K_{ij} = \frac{1}{2} \left( \dot h_{ij} - 2 D_{(i} N_{j)} \right)\,,
\end{align}
and $R[h]$ the three dimensional intrinsic curvature and the dots signalize again higher order corrections. The $c_i=c_i(t)$ are generic functions of time. The introduction of $\dot \phi$ as a generic function of time is for later convenience and can be absorbed by redefining $c_i$. The invariance of \eqref{eq:actionunitary} under the disformal transformation is trivial since the action does not depend on the lapse function. It should be noted that a priory we could also introduce terms which break the spatial covariance like $N^k N_k$. However, for simplicity, we will focus on spatial covariant gravity models. It is straightforward to check that both approaches are equivalent. By restoring the full covariance using that 
$n_\mu = \partial_\mu \phi/\sqrt{X}$, we find that 
\begin{align}\label{eq:restorecovariance}
   \sqrt{h} \dot \phi \cong {\cal V}, \qquad \dot \phi^{-1} E_{\mu\nu} \cong - {\cal B}_{\mu\nu}\quad{\rm and}\quad \qquad R[h] \cong {\cal P}^{\mu\nu} {\cal P}^{\alpha\beta} {\cal D}_{\mu\alpha\nu\beta}\,,
\end{align}
where $\cong$ means equality only in the unitary gauge. With the prescription \eqref{eq:restorecovariance} we recover the action \eqref{eq:actioncovariant} from \eqref{eq:actionunitary}.

Before going into the next section, it is instructive to look at the Hamiltonian of  \eqref{eq:actionunitary}, with $c_1=0$, but treat the scalar field $\phi$ as an independent fundamental field to understand its role in the disformal symmetry in the case of homogeneous slicing $\phi=\phi(t)$.  To simplify the calculations we split $h_{ij}$ into
\begin{align}
h_{ij}=e^{\rm 2\Psi}\Upsilon_{ij}\,,
\end{align}
where $e^{\rm 6\Psi}=\det h$, $\Upsilon^{ij}\dot\Upsilon_{ij}=0$ and $\det \Upsilon=1$ (see e.g. \cite{Domenech:2017ems}).  We also introduce for convenience a Lagrange multiplier $\lambda(\dot\phi-\chi)$ in \eqref{eq:actionunitary}. With these new variables the total Hamiltonian reads
\begin{align}\label{eq:totalH}
{\cal H}=\chi(\pi_\phi+{\cal H}_N)+N^i{\cal H}_i\,,
\end{align}
where $\pi_\phi$ is the conjugate momenta to $\phi$ and ${\cal H}_N$ and ${\cal H}_i$ are the standard Hamiltonian and momentum constraints, i.e. as if we had $N$ instead of $\dot\phi$. Concretely, ${\cal H}_N$ and ${\cal H}_i$ are respectively given by
\begin{align}
{\cal H}_N&=e^{-3\Psi}\left(\frac{1}{c_3}\pi^{ij}\pi_{ij}+\frac{\pi_\Psi^2}{12(c_3+3c_4)}\right)-e^{3\Psi}c_0-e^\Psi c_2\left(R[\Upsilon]+2D_k\Psi D^k\Psi\right)\,,\\
{\cal H}_i&=\pi_\Psi D_i\Psi -\frac{1}{3}D_i\pi_\Psi-2\Upsilon_{ij}D_k\pi^{kj}\,,
\end{align}
where $\pi^{ij}$ and $\pi_\Psi$ are respectively the conjugate momenta of $\Upsilon_{ij}$ and $\Psi$. Looking at the total Hamiltonian \eqref{eq:totalH}, we see that $\chi$ is related to the time reparametrization invariance, which is usually the role of the lapse function $N$. Noting that $\chi=\dot\phi$, we see that disformal symmetry imposes $\dot\phi$ to play the role of the lapse.
 
Taking the variation with respect to $\chi$ we find the new Hamiltonian constraint, that is
\begin{align}
\pi_\phi+{\cal H}_N=0\,.
\end{align}
In the case where all $c_i$ are constant, variation with respect to $\phi$ leads to
\begin{align}
\dot\pi_\phi=0\,,
\end{align}
which implies that ${\cal H}_N={\rm constant}$. In a Friedmann-Lemaître-Robertson-Walker (FLRW) background metric this ``integration'' constant plays the same role as mimetic dark matter, i.e. as a dust fluid with energy density decaying as $a^{-3}$. Note that something similar occurs in projectable Hořava-Lifshitz gravity \cite{Mukohyama:2009mz}. In the case where $c_i$ depends on $\phi$, we obtain instead
\begin{align}
\dot\pi_\phi+\frac{\partial {\cal H}_N}{\partial \phi}\dot\phi=0\,.
\end{align}
Unfortunately, in this general case it seems difficult to find a general solution without specifying the functional form of $c_i(\phi)$.

\subsection{Connection to Mimetic Gravity \label{subsec:mimetic}}

We proceed to show that all generic disformal invariant actions are equivalent to general mimetic gravity and viceversa. For simplicity, let us start with the latter. To see the equivalence, let us write \eqref{eq:gmatter} in terms of the metric and the scalar field. This corresponds to the degenerate disformal transformation given by \cite{Deruelle:2014zza}
\begin{align}\label{eq:mimeticg}
    \bar g_{\mu\nu} = g_{\mu\nu} + \left(\frac{1}{X} - 1\right) \nabla_\mu\phi\nabla_\nu\phi \,.
\end{align}
After a brief algebra, one finds that the metric $\bar g_{\mu\nu}$ \ref{eq:mimeticg} is invariant under generic disformal transformations \eqref{eq:disf} of $g_{\mu\nu}$. Thus, any (non disformal invariant) action or quantity expressed in terms of $g_{\mu\nu}$ through the degenerate $\bar g_{\mu\nu}$ \eqref{eq:mimeticg} must become invariant under disformal transformations of $g_{\mu\nu}$. It is also interesting to understand the meaning of \eqref{eq:mimeticg} in the uniform-$\phi$ slicing with the orthogonal vector $n_\mu$. What the transformation \eqref{eq:mimeticg} is doing is canceling the term $n_\mu n_\nu$ in the $3+1$ decomposition of $g_{\mu\nu}$ \eqref{eq:ginh} and replacing it by $\nabla_\mu\phi\nabla_\nu\phi$ in the $3+1$ decomposition of $\bar g_{\mu\nu}$, namely $\bar g_{\mu\nu}=h_{\mu\nu}-\nabla_\mu\phi\nabla_\nu\phi$. Since both $h_{\mu\nu}$ and $\nabla_\mu\phi$ are disformal invariant, the metric $\bar g_{\mu\nu}$ is disformal invariant. 
%\textcolor{blue}{From my understanding our current argumentation only covers one direction: Why a theory written in term of $\bar g_{\mu\nu}$ is disformal invariant? I think we need to provide a more detailed discussion of the other direction as well: Why all disformal invariant models can be written in terms of $\bar g_{\mu\nu}$. It should be a direct consequence of the disformal invariance but we should stress it alternatively we could provide a complete matching between both frames like in eq. 2.24. The only missing part is ${\cal D}_{\mu\nu\alpha\beta}$ in terms of $\bar R_{\mu\nu\alpha\beta}$. } \textcolor{red}{I can do that.}

Let us now show that disformal invariant actions are mimetic gravity with explicit calculations. Using \eqref{eq:mimeticg} we find that 
\begin{align}
   {\cal V} = \sqrt{- \bar g}, \qquad   {\cal P}_{\mu\nu}=\bar g_{\mu\nu} + \bar \nabla_\mu \phi \bar \nabla_\nu \phi, \qquad    {\cal N}^\nu=\bar g^{\mu\nu} \bar \nabla_\nu \phi, \\{\cal B}_{\mu\nu}=\bar \nabla_\mu \bar \nabla_\nu \phi,\qquad
   {\cal D}_{\alpha\beta\mu\nu}= \bar R_{\alpha\beta\mu\nu}-2\bar \nabla_\alpha \bar \nabla_{[\mu} \phi\nabla_{\nu]} \bar \nabla_\beta \phi\,.
\end{align}
In this way, we can write any generic disformal symmetric model in terms of the degenerate metric \eqref{eq:mimeticg}.
Since the inverse is also true, we conclude that generic disformal symmetric actions are an equivalent formulation of general mimetic gravity. At quadratic and cubic order this falls into mimetic DHOST \cite{Langlois:2018jdg}. Note that any additional term that breaks the disformal symmetry, e.g. matter fields, would take us out of mimetic gravity. To preserve the disformal symmetry, matter fields must minimally couple to $\bar g$, as we discussed in \S\ref{sec:covariant}. We will, nevertheless, show later that by relaxing the disformal symmetry we can go beyond mimetic gravity. 

Before generalizing our results, let us show that the action \eqref{eq:actioncovariant0} can also be expressed with the typical Lagrange multiplier and constraint of general mimetic gravity \cite{Golovnev:2013jxa,Langlois:2018jdg,Ganz:2018mqi}.
We first use that the general Lagrangian \eqref{eq:actioncovariant} can be expressed in terms of $\bar g_{\mu\nu}$, that is
$L(g_{\mu\nu},\phi) = L(\bar g_{\mu\nu}(g_{\mu\nu},\phi),\phi)$.
Therefore the action can be written as
\begin{align}
    S_{\rm gen.D} = \int \md^4x\, L( \bar g_{\mu\nu}(g_{\mu\nu},\phi),\phi) + \sqrt{-\bar g} \lambda_{\mu\nu} \left( \bar g^{\mu\nu} - g^{\mu\nu} + \frac{1-X}{X^2} \phi^\mu\phi^\nu \right)\,.
\end{align}
Adding the Lagrange multiplier term does not change the equations of motion since solving for $\bar g_{\mu\nu}$ fixes $\lambda_{\mu\nu}=0$.
Following \cite{Jirousek:2022jhh} we can split
\begin{align}
    \lambda_{\mu\nu} = \tilde \lambda_{\mu\nu} + \lambda \nabla_\mu\phi \nabla_\nu\phi\,,
\end{align}
with $\tilde \lambda_{\mu\nu} \phi^\mu=0$, and we obtain
\begin{align}
    S_{\rm gen.D}= \int \md^4x\, L(\bar g_{\mu\nu},\phi) + \sqrt{-\bar g} \left( \tilde \lambda_{\mu\nu} \left( \bar g^{\mu\nu} - g^{\mu\nu} \right) - \lambda \left( \bar X -1 \right) \right)\,.
\end{align}
The action does not depend anymore on $g^{\mu\nu}$ so that we can integrate it out and we are left with an action given by
\begin{align}
    S_{\rm gen.D}= \int \md^4x\, L(\bar g_{\mu\nu},\phi) - \sqrt{-\bar g} \lambda (\bar X-1)\,.
\end{align}
Now the action is clearly general mimetic gravity with the mimetic constraint, where e.g. $ L(\bar g_{\mu\nu},\phi)$ up to cubic order is a DHOST theory. We emphasize however that $ L(\bar g_{\mu\nu},\phi)$ could be any general non-disformal invariant action outside DHOST. In the most general case, this is called extended mimetic gravity in \cite{Takahashi:2017pje}.

\section{Invariance under special disformal transformations \label{sec:restricted}}

Disformal symmetry \eqref{eq:disf} with a general factor $D(x^\mu)$ leads to mimetic gravity. However, we have advanced in \S\ref{sec:generic} that there is a special disformal transformation which also leads to invariance under rescaling of the slicing. The special disformal transformation reads
\begin{align}\label{eq:specialg}
     g_{\mu\nu} \to g_{\mu\nu} + \frac{d(\phi)}{X} \nabla_\mu\phi \nabla_\nu\phi\,.
\end{align}
Such metric transformation renders the transformation of the kinetic term of $X$ as a mere field redefinition, explicitly
\begin{align}
X\to \frac{X}{1-d(\phi)}\,.
\end{align}
However, such rescaling by a function of $\phi$ only occurs for $X$ and $\sqrt{-g}$.
This means that in addition to the generic disformal invariant quantities defined in \eqref{eq:volume}-\eqref{eq:D}, we can only add
\begin{align}\label{eq:amu}
{\cal A}_\mu=-\frac{1}{2X}\left(\nabla_\mu X+\frac{\nabla_\alpha\phi\nabla^\alpha X}{X}\nabla_\mu\phi\right)=-\frac{1}{2X}P_{\mu}\,^\nu {\nabla_\nu X}\,.
\end{align}
Other terms including higher derivatives of $\phi$ lead to derivatives of $X$ after the transformation \eqref{eq:specialg} and when looking for disformal invariant tensors we go back to \eqref{eq:volume}-\eqref{eq:D}. Thus, ${\cal A}_\mu$ is the only new addition. In the action we can add terms such as $P^{\mu\nu}{\cal A}_\mu{\cal A}_\nu$, namely
\begin{align}\label{eq:specialD}
S_{\rm spec.D}=S_{\rm gen.D}+\int d^4x {\cal V}\left\{d_1(\phi) P^{\mu\nu}{\cal A}_\mu{\cal A}_\nu+\ldots\right\}\,,
\end{align}
where $S_{\rm gen.D}$ is the disformal symmetric action for generic D  \eqref{eq:actioncovariant} and the $\ldots$ now include contractions of the generic disformal invariant tensors with ${\cal A}_\mu$. 

If we relax further the condition of disformal invariance to $d={\rm constant}$ in \eqref{eq:specialg} then the term ${\nabla_\mu X}/{X}$ is invariant by itself. In that case we have that the action has additional terms given by
\begin{align}\label{eq:constd}
S_{\rm const.d}=S_{\rm spec.D}+\int d^4x {\cal V}\left\{f_0(\phi) {\cal T}+\ldots\right\}\,,
\end{align}
where we defined 
\begin{align}
\label{eq:Tcov}
{\cal T} = \frac{1}{2X} {\cal N}^\mu \nabla_\mu X=\frac{1}{2X^2} \nabla^\mu\phi \nabla_\mu X\,,
\end{align}
and the $\ldots$ include other  products of ${\cal T}$ with the other tensors \eqref{eq:volume}-\eqref{eq:D} and \eqref{eq:amu}. Let us discuss each case, i.e. the action \eqref{eq:specialD} and \eqref{eq:constd}, separately below.

\subsection{Invariance under time-dependent rescaling of the lapse}
In order to understand the transformation it is helpful to consider it again in the unitary gauge in which case the disformal transformation of the form \eqref{eq:specialg} correspond to
\begin{align}
    N \rightarrow d(t) N\,.
\end{align}
The invariant vector ${\cal A}_\mu$ \eqref{eq:amu} corresponds to the acceleration vector
\begin{align}
    ({\cal A}_u)_\mu  = a_\mu = \frac{D_\mu N}{N}\,.
\end{align}
Due to the new operator breaking the generic invariance, the action \eqref{eq:specialD} up to quadratic order does not belong anymore to the mimetic gravity class, as the whole action cannot be written in terms of $\bar g$. It is also not anymore inside DHOST but instead pertains to U-DHOST \cite{DeFelice:2018ewo} (see appendix \ref{sec:U-DHOST} for a short recap of (U)-DHOST). The theory has three degrees of freedom for a homogeneous scalar field $\phi=\phi(t)$ and outside this foliation there is an additional instantaneous mode. However, as discussed in \cite{DeFelice:2018ewo,DeFelice:2021hps}, the additional mode can be removed by imposing proper spatial boundary conditions. 

In general, the number of ${\cal A}_\mu$ will be even for each operator due to the orthogonality of ${\cal A}_\mu {\cal N}^\mu=0$. We would need to consider covariant derivatives of the other tensor like ${\cal A}_\mu {\cal P}^{\mu\nu} \nabla_\nu {\cal B}$. Therefore, without higher derivatives for linear perturbations around a homogeneous background like FLRW such operators can be neglected since they vanish identically. At the background level this follows directly from requiring a homogeneous background and at linear order we obtain terms like
\begin{align}
    S^{(2)}_{\rm spec.D}= S^{(2)}_{\rm gen.D} + \int \md^3x\,\md t\, \bar f^{ij} \partial_i \delta N \partial_j \delta N\,,
\end{align}
where $\bar f^{ij}$ purely depends on the background quantities. Deriving the equations of motion for $\delta N$ leads to $\bar f^{ij} \partial_i \partial_j \delta N =0$ and, therefore, by imposing proper boundary conditions we can set $\delta N=0$. Therefore, for homogeneous backgrounds at the linear level the perturbations are equivalent to the mimetic gravity model. 

\subsection{Invariance under a constant rescaling of the lapse}

In the case where $d(\phi)=\mathrm{const}$ the new term \eqref{eq:Tcov} in the unitary gauge can be expressed as
\begin{align}
    {\cal T} = - \frac{1}{\dot \phi}\left\{\frac{\md}{\md t} \ln\left( \frac{\dot \phi}{N} \right) -  N^k D_k \ln\left( \frac{\dot \phi}{N} \right) \right\}\,.
\end{align}
In the unitary gauge it is easy to see that it will lead to an additional degree of freedom due to the time derivative of the lapse function. In the covariant formulation we can check that introducing these terms will break the degeneracy conditions of DHOST and U-DHOST. The additional degree of freedom is what it is commonly called the Ostrogradsky ghost. However, in covariant theories the analysis of Ostrogradsky ghost instabilities is a bit more subtle since the theory has a first class Hamiltonian constraint \cite{Ganz:2020skf}. In the next two sections, we discuss that for linear perturbations around FLRW the perturbations can be stable despite the presence of the Ostrogradsky ghost.

\section{Cosmology and cosmological perturbations\label{sec:cosmo}}
Let us discuss the general case of linear perturbations around FLRW. We consider the action up to the quadratic order given by
\begin{align}\label{eq:actionweuse}
    S_{\rm constant.d} = \int \md^4x\, {\cal V}  \Big[ c_0  + c_2 {\cal P}^{\mu\alpha} {\cal P}^{\nu\beta} {\cal D}_{\alpha\beta\mu\nu} + c_3 {\cal B}^2 + c_4 {\cal B}_{\mu\nu} {\cal B}^{\mu\nu} + d_1 {\cal A}_\mu {\cal A}^\mu + f_1 {\cal T}^2 - f_2 {\cal T} {\cal B} \Big]\,.
\end{align}
For simplicity, we will assume that the coefficients $c_i$, $d_1$ and $f_j$ are constant and $f_2=0$. We will discuss the impact of a kinetic coupling between both degrees of freedom in the next section. We did not include terms linear in ${\cal B}$ or ${\cal T}$ since for constant coefficients they are total derivatives.

\subsection{Background level}
Let us first consider the background level, where the line element reads 
\begin{align}\label{eq:dsbg}
    \md s^2 = - N^2 \md t^2 + a^2 \delta_{ij} \md x^i \md x^j\,.
\end{align}
The Lagrangian of \eqref{eq:actionweuse} for \eqref{eq:dsbg} is then given by
\begin{align}\label{eq:Lweuse}
    L = a^3 q N \left[ c_0 +  3 \frac{(c_3 + 3 c_4)}{q^2 N^2} H^2 + \frac{f_1}{q^4 N^2} \dot q^2   \right] + \pi (\dot \phi - q N)\,,
\end{align}
where we have introduced the new variable $q = \dot \phi/N$, $\pi$ is the associated canonical conjugate momenta to $\phi$ and $H=\dot a/a$ is the Hubble parameter.
The other momenta are given by
\begin{align}
    \frac{p_q}{a^3} =  \frac{2 f_1}{q^3 N} \dot q, \qquad \frac{p_a}{a^3} = \frac{6 (c_3 + 3c_4)}{q N a} H\qquad{\rm and} \qquad p_N \approx0\,,
\end{align}
where the last is a primary constraint and $\approx$ denotes weak equality on the constrained hypersurface. 
The total Hamiltonian from \eqref{eq:Lweuse} is given by
\begin{align}
    H_T = N q H_0 + u_1 p_N = N q \left[ \frac{p_a^2 a^2}{12 a^3 (c_3 + 3 c_4)} + \frac{p_q^2 q^2}{4 a^3 f_1} - a^3 c_0 + \pi \right] + u_1 p_N\,.
\end{align}
The conservation of the primary constraint leads to the first class Hamiltonian constraint as expected due to the time reparametrization invariance
\begin{align}
    H^\prime = q H_0 \approx 0\,,
\end{align}
which leads to $q\approx 0$ or $H_0\approx0$. The former is trivial, so let us focus on the latter case. The Hamiltonian constraint contains as expected a term linear in the momentum due to the higher derivative term. Despite the presence of a linear term in the momentum, it can be shown that the system is stable if $c_3+3c_4>0$, $f_1>0$ and $c_0<0$ \cite{Ganz:2020skf}. The main reason is that $\pi$ is the momentum of the parametrized time $\phi$. 

To explicitly show that the system is stable, we fix the time reparametrization invariance from the start by choosing $\dot \phi=1$. In other words, since we always have $\dot\phi dt$ in the action we can set $\phi=t$ without loss of generality. In that case the scalar fixed Lagrangian is given by
\begin{align}\label{eq:lgf}
    L_{sf} = a^3 \left[ c_0 + 3 (c_3+3c_4) H^2 + f_1 \frac{\dot N^2}{N^2} \right]\,,
\end{align}
where the subscript ``sf'' refers to ``scalar fixed''. From \eqref{eq:lgf} we find that
\begin{align}
    H_{gf} = \frac{p_a^2 a^2}{12 a^3 (c_3+3 c_4) } + \frac{p_N^2 N^2}{4 a^3 f_1} -a^3 c_0\,,
\end{align}
which is obviously stable for an appropriate choice of $c_i$ so that the Hamiltonian is bounded from below. This is not in contradiction with the Ostrogradski ghost theorem because of the presence of a first class Hamiltonian constraint. In passing, we note that the background equations of motion in the case where $f_1=0$ lead to $\ddot\phi=0$ and, therefore, $\dot\phi={\rm constant}$, which is consistent with our choice $\dot\phi=1$. This is what happens in mimetic gravity if $t$ is the cosmic time, i.e. if $N=1$ then $\dot\phi=1$.

\subsection{Linear level}
For the linear perturbations we use the unitary gauge (that is uniform-$\phi$ and $\dot\phi=1$) where the metric components read 
\begin{align}
    h_{ij} = a^2 e^{2\zeta} \left(e^{h}\right)_{ij}, \qquad N =  N_0 + \delta N, \qquad N_k= \partial_k \beta\,.
\end{align}
From now on we will abuse notation and we drop the subscript zero for the background part of the lapse function. Let us consider first tensor perturbations. In that case, the second order action is given by
\begin{align}\label{eq:ST2}
    S_T^{(2)} = \frac{1}{4} \int \md^3 k\,\md t\, a^3 \dot \phi \left[ \frac{c_3}{\dot \phi^2} \dot h_{ij}^2 - c_2 \frac{k^2}{a^2} h_{ij}^2 \right]\,.
\end{align}
We see that \eqref{eq:ST2} takes the usual form, but with $\dot\phi$ instead of $N$, as if we used the constraint $N=\dot\phi$ in general mimetic gravity. Second, variation with respect to the non-dynamical shift $\beta$ yields
\begin{align}
\label{eq:constraint_beta}
    (c_3+c_4) \frac{k^2}{a^2} \beta =  f_1 \frac{\dot N}{N} \frac{\delta N}{N} - (c_3+ 3 c_4) \dot \zeta\,.
\end{align}
We will assume for the moment that $c_3\neq - c_4$, and come back to it later. Plugging $\beta$ back into the second order action for scalar perturbations yields, after simplifications, 
\begin{align}\label{eq:S2dN}
    S^{(2)} = \int \md^3k\,\md t\, a^3 \Big[ & \frac{2 c_3 (c_3 + 3 c_4)}{c_3 + c_4} \dot \zeta^2 + 2 c_2 \frac{k^2}{a^2} \zeta^2 + f_1\frac{\delta \dot N^2 }{N^2} 
    +  \left( d_1 \frac{k^2}{a^2} - f_1 \frac{c_3 + c_4 + f_1}{(c_3 + c_4)} \frac{\dot N^2}{N^2}\right) \frac{\delta N^2}{N^2}\nonumber \\&-\frac{2 c_3 f_1 \dot N^2}{(c_3+c_4) N^2}  \zeta \frac{\delta  N}{N}  + \frac{2 c_3 f_1 \dot N}{(c_3+c_4) N} \left( \zeta \frac{\delta \dot N}{N}-\dot \zeta \frac{\delta N}{N}\right) \Big]\,.
\end{align}
It is interesting to note that we can tune the extra degree of freedom %, second line in \eqref{eq:S2dN},
to be stable (no gradient or ghost instability) by choosing $f_1>0$ and $d_1<0$. However, the first two terms in the first line of \eqref{eq:S2dN} remains the same as the mimetic degree of freedom, which is in general unstable. If we choose $c_2 <0$ then the gradient term for the tensor modes \eqref{eq:ST2} has the wrong sign. This is a known problem of general mimetic gravity (see for instance \cite{Langlois:2018jdg,Firouzjahi:2017txv,Hirano:2017zox,Takahashi:2017pje,Zheng:2017qfs,Gorji:2017cai}) and, therefore, it is not connected to the additional degree of freedom due to the dynamical lapse. We also note that the mixing terms in the second line are only relevant for tachyonic like instabilities and are not linked to gradient or ghost instabilities \cite{DeFelice:2016ucp}. We also note that the case $c_2=0$ corresponds to no spatial Ricci scalar in the action and leads to non-propagating $\zeta$ and $h_{ij}$. From now on we always consider $c_2>0$.

Let us consider the special case $c_3=-c_4$ in more detail in which the mimetic fluid behaves exactly as dust with vanishing sound speed, that is as mimetic dark matter. Considering first $d_1=f_1=0$, $c_4=-c_3=c_2=1/2$  we have the original mimetic matter model in which case it has been shown that the model suffers under a ghost or tachyon instability \cite{Ganz:2018mqi}.\footnote{Despite the presence of $c_2 k^2 \zeta^2$ in the second order action the equation of motion do not have any scale dependency (vanishing sound speed) due to $\dot \zeta=0$ coming from the momentum constraint. Therefore, it is not possible to distinguish between the ghost or tachyon instability since they are related via a canonical transformation.} Including the dynamical lapse function, i.e. $d_1,f_1\neq0$,  the momentum constraint \eqref{eq:constraint_beta} fixes
\begin{align}
    \frac{\delta N}{N} = \frac{2 c_4 N}{f_1 \dot N} \dot \zeta\,.
\end{align}
Substituting it back into the action leads to higher order time derivatives in the action resulting in a Ostrograski ghost instability
\begin{align}
    S^{(2)}({c_3=-c_4}) = \int \md^3k\,\md t\,a^3 \Big[& \frac{4 c_4^2  N^2 }{f_1 \dot N^2} \ddot \zeta^2 + \frac{c_4}{f_1^2 \dot N^2 } \left( -3 f_1 (c_0 N^2 + 18 c_4 N^2 H^2 + 3 f_1 \dot N^2) + 4 c_4 d_1 \frac{k^2}{a^2} \right) \dot \zeta^2 \nonumber \\
    & + 2 c_2 \frac{k^2}{a^2} \zeta^2 \Big]\,.
\end{align}
This is similar to the case of mimetic gravity plus an additional normal matter fluid \cite{Langlois:2018jdg,Takahashi:2017pje}. However, in that case the instability is still directly linked to the dust itself \cite{Ganz:2018mqi}. Due to the same structure of the action we expect that the analysis of \cite{Ganz:2018mqi} also applies to our model and we will not consider it further.

\section{Higher curvature term\label{sec:highercurvature}}

We have seen that the special disformal symmetry \eqref{eq:specialg} with a constant coefficient, in a perturbed FLRW universe, leads to a healthy extra degree of freedom. Unfortunately, due to the mimetic-like nature of the action, we found that at the lowest order in the disformal invariant operators the second order action perturbations is accompanied by some kind of instability. 
If the mimetic-like degree of freedom behaves like dust there is a tachyonic or ghost instability and if it has a non-vanishing sound speed there is a ghost or gradient instability. Nevertheless, it has been shown that in order to obtain stable linear perturbations one could add higher curvature terms, i.e. couplings of the form $E R[h]$ \cite{Gorji:2017cai,Zheng:2017qfs,Hirano:2017zox} (but see e.g.~\cite{Babichev:2017lrx} for other potential solutions involving additional scalar fields). In our disformal symmetric formulation we are allowed to include higher order operators by contractions of disformal invariant tensors. At the covariant level the relevant terms are
\begin{align}
    {\cal D}_{\mu\nu\alpha\beta} {\cal P}^{\mu\alpha} {\cal B}^{\nu\beta}, \quad {\cal D}_{\mu\nu\alpha\beta} {\cal P}^{\mu\alpha} {\cal P}^{\nu\beta} {\cal B}, \quad {\cal D}_{\mu\nu\alpha\beta} {\cal P}^{\mu\alpha} {\cal P}^{\nu\beta} {\cal T} , \quad {\cal D}_{\mu\nu\alpha\beta} {\cal P}^{\mu\alpha} {\cal N}^{\nu} {\cal N}^\beta {\cal B}, \quad {\cal D}_{\mu\nu\alpha\beta} {\cal P}^{\mu\alpha} {\cal N}^{\nu} {\cal N}^\beta {\cal T}\,.
\end{align}
However, these terms normally add even more degrees of freedom since they contain time derivatives of the extrinsic curvature \cite{Ganz:2019vre,Zheng:2018cuc}. In fact, there are two components of the extrinsic curvature which acquire time derivatives, namely 
\begin{align}
    \frac{D^i \phi D^j \phi}{D_k \phi D^k\phi} E_{ij}, \qquad \left( h^{ij} - \frac{D^i \phi D^j\phi}{D_k \phi D^k \phi}\right) E_{ij}\,.
\end{align}
That being said, the first term vanishes trivially in the unitary gauge and we can impose additional degeneracy conditions to get rid of the second one, which leads to a generalized class of U-DHOST. In our case, we are interested in the three different operators:
\begin{align}
    \frac{E}{\dot \phi} R[h], \qquad \frac{E_{ij}}{\dot \phi} R^{ij}[h]\qquad {\rm and} \qquad {\cal T} R[h]\,,
\end{align}
where the last term is only valid for the relaxed disformal transformation $N \rightarrow d \times N$ with $d=\mathrm{const}$. Therefore, there are two additional degeneracy conditions between the covariant operators to remove the time derivatives of the extrinsic curvature in the unitary gauge. Let us emphasize that the higher curvature terms are only needed to stabilize the mimetic-like degree of freedom.

\subsection{Stable model}
Let us show that by including the coupling between $E$ and $R[h]$ we get stable linear perturbations. To simplify the calculations we will only consider the quadratic terms and the cubic terms which impact the gradient terms for the scalar sector. In the unitary gauge this corresponds to
\begin{align}\label{eq:shigherD}
    S_{\rm const.d} =& \int \md^3x\,\md t\,\sqrt{h} \dot \phi \Big[ c_0 + c_2 R[h] + \frac{c_3}{\dot \phi^2} E_{ij} E^{ij} + \frac{c_4}{\dot \phi^2} E^2 + \frac{c_5}{\dot \phi} E R[h] + \frac{c_6}{\dot \phi} R_{ij}[h] E^{ij} + d_1 a_i a^i  + \frac{d_2}{\dot \phi} a_j a^j E \nonumber \\
    & + \frac{d_3}{\dot \phi} a^i a^j E_{ij}  + f_1 {\cal T}^2 + f_2 \frac{E}{\dot \phi} {\cal T} + f_3 {\cal T} R[h] \Big]\,.
\end{align}
For simplicity, we take $\phi=t$ from now on. Note that if we are only interested in the linear perturbations around FLRW we can absorb the effect of $d_3$ by redefining $d_2$ and, similarly, $c_6$ into $c_2$ and $c_5$. Therefore, we will set $c_6=d_3=0$ in what follows without loss of generality. We proceed to derive the equations for the background and linear perturbations.

First, at the background level we find
\begin{align}
    & \dot H = \frac{- 3 (2 c_3 f_1+ 6 c_4 f_1 - 3 f_2^2)H^2 + 6 f_1 f_2 H H_N+ 2 f_1 (c_0 + f_1 H_N^2)}{4 c_3 f_1 + 12 c_4 f_1 - 3 f_2^2} \,, \\
    & \dot H_N =- 3 \frac{3 (c_3 + 3 c_4) f_2 H^2+ 4 (c_3 + 3 c_4) f_1 H H_N+ f_2 (c_0 + f_1 H_N^2)}{4 c_3 f_1 +12 c_4 f_1 - 3 f_2^2}\,,
\end{align}
where we defined $H_N = \dot N/N$. For linear perturbations of tensor modes, we find that the second order action reads
\begin{align}
    S^{(2)}_T = \frac{1}{4} \int \md^3 k\md t\, a^3 \left( c_3 \dot h_{ij}^2 - ( c_2 + 3 c_5  H + f_3 H_N ) \frac{k^2}{a^2} h_{ij}^2\right)\,.
\end{align}
We see that to have stable perturbations we must require
\begin{align}
    c_3>0\quad {\rm and}\quad c_2 + 3 c_5 H + f_3 H_N >0\,.
\end{align}

Let us turn now to the scalar sector, where we again find two degrees of freedom. Variation of \eqref{eq:shigherD} with respect to the shift $\beta$ leads to
\begin{align}
    2 N  (c_3 + c_4) \frac{k^2}{a^2}\beta =& (3 f_2 H + (2 f_1+f_2) H_N) \delta N - 4 N c_5 \frac{k^2}{a^2} \zeta  - f_2  \delta \dot N - 2  (c_3 + 3 c_4 ) N \dot \zeta  \,.
\end{align}
This time, however, after using the solution for $\beta$, the resulting second order action contains a quadratic kinetic mixing between $\dot\zeta$ and $\delta\dot N$ proportional to $f_2$.  To diagonalize the kinetic terms we introduce 
\begin{align}
    \psi_1 = \zeta + \frac{f_2}{2 c_3 + 6 c_4} \frac{\delta N}{N}\qquad{\rm and} \qquad \psi_2=\frac{\delta N}{N}\,.
\end{align}
After some partial integrations the second order action for scalar perturbations takes the form
\begin{align}\label{eq:diagonal}
    S^{(2)}=\int \md^3k \md t\, a^3 \left(  K_1 \dot \psi_1^2 + K_2 \dot \psi_2^2 + B (\dot \psi_1 \psi_2 - \psi_1 \dot \psi_2) -  \psi_j G_{ij} \psi_i  \right)\,,
\end{align}
where we defined
\begin{align}
    &K_1 = \frac{2 c_3 (c_3 + 3 c_4) }{c_3 + c_4}, \quad K_2 = \frac{4 f_1 (c_3 + 3 c_4) - 3 f_2^2}{4 (c_3 + 3 c_4)}\,,  \\
    & B = \left( \frac{c_5 f_2 }{c_3 + c_4} - 2f_3 \right) \frac{k^2}{a^2} - \frac{c_3}{(c_3 + c_4)} \left(3 f_2 H + 2 f_1 H_N \right)\,, \\
    &G_{11} = \frac{4 c_5 }{c_3 + c_4} \frac{k^4}{a^4} - 2 \frac{(c_3 + 3 c_4) c_5 H + (c_3 + c_4) (c_2 + f_3 H_N)}{c_3 + c_4} \frac{k^2}{a^2}\,, \\
    & G_{22} = \frac{c_5^2 f_2^2}{(c_3 + c_4) (c_3 + 3 c_4)^2 } \frac{k^4}{a^4} + C_{22}^2 \frac{k^2}{a^2} + \frac{(3 f_2 H+ 2 f_1 H_N)^2}{4(c_3+c_4)}\,,\\
    & G_{12} = - \frac{2 c_5^2 f_2 }{(c_3 + c_4) (c_3 + 3 c_4) } \frac{k^4}{a^4} + C_{12}^2 \frac{k^2}{a^2}\,.
\end{align}
For completeness, we also show the explicit expressions for $C_{22}^2$ and $C_{12}^2$, which are respectively given by
\begin{align}
    C_{22}^2 =& -\frac{1}{2 (c_3 + c_4) (c_3 + 3c_4)^2 }\Big[  2 (c_3+c_4) (c_3+3c_4)^2 d_1  +f_2H_N (f_2 f_3 (c_3 + c_4) -4c_5 f_1 (c_3 + 3c_4)  ) \nonumber \\
    & +  c_2f^2_2 (c_3 + c_4)+ 2H (c_3+3c_4)  ((c_3+c_4)(3(c_3+3c_4)d_2+f_2f_3)-3c_5f^2_2)\Big]\,, \\
    C^2_{12}=& \frac{2 c_2 f_2 (c_3+c_4) + (c_3+3 c_4) (2 (c_3+c_4) f_3-5 c_5 f_2) H+ 2 (f_2 f_3 (c_3+c_4)-2  c_5 f_1 (c_3+3c_4)  ) H_N}{2 (c_3 + c_4) (c_3+3 c_4)} \,.
\end{align}

By looking at \eqref{eq:diagonal}, we find that no ghost conditions are given by $K_1 >0$ and $K_2>0$ which can be fulfilled by choosing proper parameters for $c_i$ and $f_i$. The coefficient $B$ again does not impact the ghost instability due to the antisymmetric structure \cite{DeFelice:2016ucp}. As it is clear from \eqref{eq:diagonal}, the gradient matrix $G_{ij}$ is not diagonal in general due to the mixing term proportional to $f_2$ and $f_3$. Furthermore, it contains terms up to the order $k^4$ due to the term proportional to $c_5$ as occurs in general mimetic gravity with higher order corrections. 

We can find the stability conditions for the gradient part as follows \cite{DeFelice:2016ucp}. We first derive the coupled equations of motion for $\psi_1$ and $\psi_2$. We then focus on the limit $k\gg aH$ and neglect any time dependence of the coefficients in \eqref{eq:diagonal}. And, we derive  dispersion relation in the high momenta limit by first using the ansatz $\psi_1,\psi_2\sim e^{-i\omega t}$ and requiring that the system is degenerate. The determinant of the coupled system vanishes when
\begin{align}
     K_1 K_2 \omega^4 -  \omega^2 (G_{11} K_2 + K_1 G_{22} + B^2) -  G_{12}^2 +  G_{11} G_{22}\simeq 0\,.
\end{align}
Noting that $- G_{12}^2+  G_{11} G_{22} = \mathcal{O}(k^6)$ we see that there are two solutions at leading order in the high $k$ limit, namely $\omega = C_{\omega 1} k^2/a^2$ and $\omega = C_{\omega 2}  k/a$. After some algebra, we find
\begin{align}
    C_{\omega 1} ^2 =& 8 \frac{c_5^2 f_1 - c_5 f_2 f_3 + (c_3 + c_4) f_3^2}{c_3 (4f_1 (c_3 + 3 c_4) - 3 f_2^2 )}\,, \\
    C_{\omega 2} ^2 =&  -  \frac{c_5^2 (d_1 + 3 d_2 H ) }{(c_5^2 f_1 - c_5 f_2 f_3 + (c_3 + c_4) f_3^2) N^2}\,.
\end{align}
By choosing proper coefficients both $C_{\omega 1}^2$ and $C_{\omega 2}^2$ can be positive definite. As an example, the stability conditions can be fulfilled for $c_i>0$, $f_1 (c_3+3c_4) > 3 f_2^2$, $f_2 f_3 <0$ and $d_i < 0$ (assuming an expanding spacetime, namely $H>0$). We conclude that by adding the coupling between the extrinsic and intrinsic curvature $E R[h]$ we can obtain stable linear perturbations (no ghost or ultraviolet gradient instabilities) despite the presence of the additional degree of freedom due to the higher derivative terms. 

\section{Conclusions \label{sec:conclusions}}

We investigated the meaning and implications of pure disformal symmetry in the Universe. We built general disformal invariant tensors involving first and second derivatives of a scalar field as well as the curvature tensor, given by equations \eqref{eq:volume}--\eqref{eq:D}. We showed that requiring generic disformal invariance of the action \eqref{eq:actioncovariant}, that is invariance under a pure disformal transformation with an arbitrary coefficient, is equivalent to a lapse-less EFT-like action in the unitary gauge \eqref{eq:actionunitary}. Up to cubic order, disformal symmetric theories can be mapped DHOST and U-DHOST (see appendix \ref{sec:U-DHOST}). But, disformal symmetry is more general as it allows one to write higher order operators outside of (U-)DHOST.

We then proved that the generic disformal symmetric action \eqref{eq:actioncovariant} is an equivalent formulation to general mimetic gravity \cite{Chamseddine:2013kea,Langlois:2018jdg,Takahashi:2017pje}. This means that, in general, all disformal symmetric theories are equivalent to general mimetic gravity and viceversa. In contrast, not all conformal invariant models are related to mimetic gravity. To understand the role of $\phi$ in disformal symmetric theories we have computed the Hamiltonian in the uniform-$\phi$ slicing \eqref{eq:actionunitary}. This showed that the role of the lapse as time reparametrization parameter has been replaced by the time derivative of the scalar field, without the need of invoking any constraint as usually done in mimetic gravity. Thus, in disformal symmetric theories, one may say that the lapse function has been completely replaced by a fundamental scalar field. We also find that the Hamiltonian constraint of disformal symmetric theories leads to standard Hamiltonian constraint (i.e. if we still had the lapse in the action) which does not vanish but it is proportional to the momentum of the scalar field. In the simplest case, the standard Hamiltonian constraint is equal to a constant which leads to a (mimetic) dark matter degree of freedom, in a FLRW background.

We then relaxed the generic disformal symmetry to the case of field-dependent-only or constant disformal coefficients. This enlarged the number of terms allowed in the action, respectively \eqref{eq:specialD} and \eqref{eq:constd}. In the case of constant disformal coefficient,  the special disformal symmetric action \eqref{eq:constd} contains a new degree of freedom associated to the time derivative of the lapse. The resulting theory enters in a subclass of spatially covariant gravity with dynamical lapse \cite{Gao:2018znj,Lin:2020nro}. And, in general, such extra degree of freedom is expected to be an Ostrogradski ghost. However, by analyzing linear cosmological perturbations we showed that the system is stable, except for the usual instability associated to the mimetic degree of freedom \cite{Langlois:2018jdg,Firouzjahi:2017txv,Hirano:2017zox,Takahashi:2017pje,Zheng:2017qfs,Gorji:2017cai}. We then showed that by including higher curvature terms allowed by the special disformal symmetry one can render the system stable.  

Our work can be extended in several ways. First, we restricted our analysis for simplicity to the quadratic terms in the covariant action \eqref{sec:covariant}. However, one could do a general analysis including higher order operators as in \cite{Takahashi:2017pje}. Second, it would be interesting to generalize the pure disformal symmetry to general disformal symmetry including the conformal factor. We expect that in this case the form of the resulting theory will be more constrained by the additional requirement of conformal symmetry. Whether this case also belongs to mimetic gravity is not straightforward to us. It would also be interesting to study the implications of disformal symmetry in vector-tensor theories of gravity, also known as generalized Proca theories \cite{Heisenberg:2014rta,Kimura:2016rzw,Heisenberg:2016eld,Papadopoulos:2017xxx,Domenech:2018vqj}. One possibility is that this case is related to mimetic gravity in a vector-tensor theory as in \cite{Chaichian:2014qba,Jirousek:2022kli}.

Lastly, the disformal invariant tensors we provided have applications beyond disformal invariant actions. In particular, we propose an alternative way to build disformal transformations with higher derivatives to those studied in  \cite{Babichev:2021bim,Takahashi:2021ttd,Takahashi:2023jro}. By allowing the disformal factor $D$ to depend on disformal invariant tensors, namely
\begin{align}
    \bar g_{\mu\nu} = g_{\mu\nu} + D(\phi,{\cal N}^\mu,{\cal P}^{\mu\nu},{\cal B}_{\mu\nu},{\cal D}_{\mu\nu\alpha\beta}) \nabla_\mu \phi \nabla_\nu \phi\,,
\end{align}
the metric transformation is trivially invertible and, after performing the transformation, it will lead to new theories. However, one should also take into account the matter coupling as, e.g., it may lead to ghosts in the generalized disformal transformations of \cite{Naruko:2022vuh,Takahashi:2022mew,Ikeda:2023ntu}.  As another example, disformal invariant tensors might help in constructing the notion of disformally flat spacetimes, as in \cite{Domenech:2019syf} and possible singularity-free frames for black hole spacetimes. These interesting directions are beyond the scope of this paper and we leave them for future work.

\section*{Acknowledgments}

We would like to thank A.~de Felice, A.~Naruko, M.~Minamitsuji and A.~Vikman for useful comments and discussions. G.D. is supported by the DFG under the Emmy-Noether program grant no. DO 2574/1-1, project number 496592360. A.G. receives support by the grant No. UMO-2021/40/C/ST9/00015 from the National Science Centre, Poland. Calculations of the disformal invariant tensors were cross-checked with the \textsc{Mathematica} package \textsc{xAct} (\url{www.xact.es}).

\appendix

\section{Disformal transformation formulas and invariant tensors\label{sec:disfforumlas}}

In this appendix we provide the general formulas for pure disformal transformation that are used in the main text. We always use normalized (anti)symmetrization. 

First, we define the metric transformation. A pure disformal transformation is given by
\begin{align}
\bar g_{\mu\nu}=g_{\mu\nu}+ D \partial_\mu\phi\partial_\nu\phi \,.
\end{align}
The inverse metric reads
\begin{align}
\bar {g}^{\mu\nu}=g^{\mu\nu}-\frac{D}{1-DX}\nabla^\mu\phi\nabla^\mu\phi\,,
\end{align}
where we defined $X= - g^{\mu\nu}\partial_\mu \phi \partial_\nu \phi $. We now give the transformation rules.

We have that $\bar X$ transforms as
\begin{align}
    \bar X = \frac{X}{1- DX}\,.
\end{align}
The transformation of the second order derivative term, i.e. $\bar \nabla_\mu \bar \nabla_\nu \phi$, can be expressed as
\begin{align}
    \bar \nabla_\mu \bar \nabla_\nu \phi = \nabla_\mu \nabla_\nu \phi - {\cal K}^\lambda_{\mu\nu} \nabla_\lambda \phi\,,
\end{align}
where 
\begin{align}
{\cal K}^\lambda_{\mu\nu}&=\frac{D}{1-DX}\nabla^\lambda\phi\left(\nabla_\mu\nabla_\nu\phi+\nabla_{(\mu}\phi\nabla_{\nu)}\ln D+\frac{1}{2}\nabla^\rho\phi\nabla_\rho D\nabla_\mu\phi\nabla_\nu\phi\right)
    -\frac{1}{2}\nabla^\lambda D\nabla_\mu\phi\nabla_\nu\phi\,.
\end{align}
With the formula above, we find that
\begin{align}
\bar \nabla_\mu \bar \nabla_\nu \phi = \frac{1}{1-DX}\left(\nabla_\mu \nabla_\nu \phi+X\nabla_{(\mu}\phi \nabla_{\nu)}D+\frac{1}{2}\nabla_{\alpha}\phi \nabla^{\alpha}D\nabla_\mu\phi \nabla_\nu \phi\right)\,.
\end{align}
After a longer computation, one can also show that 
\begin{align}
\bar{ R}_{\alpha \beta \mu  \nu  }=&{ R}_{\alpha \beta \mu  \nu  }+2\frac{D}{1-DX}\nabla_\alpha\nabla_{[\mu  }\phi\nabla_{\nu  ]}\nabla_\beta \phi-\frac{X}{1-DX}\nabla_{[\alpha}\phi\nabla_{\beta ]}D\nabla_{[\mu  }\phi\nabla_{\nu  ]}D+2\nabla_{[\alpha}\phi\nabla_{\beta ]}\nabla_{[\mu  }D\nabla_{\nu  ]}\phi\nonumber\\&
-2\frac{D}{1-DX}\nabla_e\phi\nabla^e D\nabla_{[\alpha}\phi\nabla_{\beta ]}\nabla_{[\mu  }\phi\nabla_{\nu  ]}\phi-2\frac{1}{1-DX}\left(\nabla_{[\alpha}\phi\nabla_{\beta ]}\nabla_{[\mu  }\phi\nabla_{\nu  ]}D+\nabla_{[\mu  }\phi\nabla_{\nu]}\nabla_{[\alpha}\phi\nabla_{\beta ]}D\right)\,.
\end{align}
After contraction with $\bar g^{\mu\nu}$ we find that
\begin{align}
	\begin{split}
		\bar{R}&={ R}-\frac{D}{1-DX}\left((\Box\phi)^2-\nabla_\mu\nabla_\nu\phi\nabla^\mu\nabla^\nu\phi
		+\nabla^\mu \ln D\left(\frac{1}{2}\nabla_\mu X+\nabla_\mu\phi \Box\phi\right)\right)\\&
		+\frac{1}{\sqrt{1-DX}}\nabla^\mu\left(\frac{D}{\sqrt{1-DX}}\left(\nabla_\mu\phi\nabla^\nu\phi\nabla_\nu \ln D+X\nabla_\mu \ln D+\nabla_\mu X+2\Box\phi\nabla_\mu\phi\right)\right)\,.
	\end{split}
\end{align}

To see the impact of the disformal transformation it is useful to consider it in the unitary gauge $\phi=\phi(t)$ or $n_\mu = - \partial_\mu \phi/ \sqrt{X}$. 
Using the standard ADM decomposition the purely disformal transformation only impacts the lapse function
\begin{align}
    \tilde N = N \Phi(t,N) \equiv N \sqrt{ 1 + \frac{\dot \phi^2}{N^2} D(t, N)}
\end{align}
and the spatial metric and the shift vector remain unchanged $\tilde h_{ij} = h_{ij}$ and $\tilde N^k = N^k $.

\subsection{Disformal invariant tensors}

With the transformation rules shown above we find the basic disformal invariant tensors.\\

\noindent\textbf{Disformal volume element:}\\

\begin{align}
{\cal V}=\sqrt{-g \,X}\,.
\end{align}

\noindent\textbf{Disformal projector:}\\
\begin{align}
{\cal P}_{\mu\nu}= g_{\mu\nu}+\frac{1}{X}\nabla_\mu\phi\nabla_\nu\phi \quad{\rm and}\quad 
{\cal P}^{\mu\nu}= g^{\mu\nu}+\frac{1}{X}\nabla^\mu\phi\nabla^\nu\phi \,.
\end{align}
\noindent\textbf{Disformal upper first derivative:}\\
\begin{align}
{\cal N}^\mu=& \frac{1}{{X}}\nabla^\mu\phi \,.
\end{align}

\noindent\textbf{Disformal second derivative:}\\
\begin{align}
{\cal B}_{\mu\nu}\equiv&\frac{1}{X}\left(\nabla_\mu\nabla_\nu\phi-\frac{1}{X}\nabla_{(\mu}X\nabla_{\nu)}\phi-\frac{1}{2X^2}\nabla_\alpha \phi\nabla^\alpha X\nabla_\mu\phi\nabla_\nu\phi \right)\,.
\end{align}
\noindent\textbf{Disformal curvature tensor:}\\
\begin{align}
{\cal D}_{\alpha\beta\mu\nu}\equiv &R_{\alpha\beta\mu\nu}-\frac{2}{X}\nabla_{\mu}\nabla_{[\alpha}\phi\nabla_{\beta]}\nabla_{\nu}\phi-\frac{2}{X^2}
\nabla_{[\alpha}\phi\nabla_{\beta]}\nabla_{[\mu}X\nabla_{\nu]}\phi
-\frac{4}{X^3}\nabla_{[\alpha}\phi\nabla_{\beta]}X\nabla_{[\mu}\phi\nabla_{\nu]}X\nonumber\\&-\frac{2\nabla_\mu\phi\nabla^\mu X}{X^3}
\nabla_{[\alpha}\phi\nabla_{\beta]}\nabla_{[\mu}\phi\nabla_{\nu]}\phi\,.
\end{align}
\\

\noindent\textbf{Example of contractions:} Here we give the simplest expressions after contraction of the above disformal tensors. First, we have that
\begin{align}
{\cal P}^{\mu\nu}\,{\cal B}_{\mu\nu}=\frac{1}{X}\left(\Box\phi-\frac{1}{2X}\nabla_\mu X\nabla^\mu \phi\right)\,.
\end{align}
It can also be shown that
\begin{align}
{\cal P}^{\alpha\mu}{\cal P}^{\beta\nu}{\cal B}_{\alpha\beta}{\cal B}_{\mu\nu}=\frac{1}{X^2}\left(\nabla_\mu\nabla_\nu\phi\nabla^\mu\nabla^\nu\phi+\frac{1}{2X}\nabla_{\mu}X\nabla^{\mu}X+\frac{1}{4X^2}\left(\nabla_\mu \phi\nabla^\mu X\right)^2\right)\,.
\end{align}
Lastly, the two possible contractions involving the curvature yields
\begin{align}
{\cal P}^{\alpha\mu}{\cal P}^{\beta\nu}{\cal D}_{\alpha\beta\mu\nu}=R+\frac{1}{X}\left((\Box\phi)^2-\nabla_\mu\nabla_\nu\phi\nabla^\mu\nabla^\nu\phi\right)-\frac{2}{\sqrt{X}}\nabla^\mu \left(\frac{1}{\sqrt{X}}\left(\Box\phi\nabla_\mu\phi+\frac{1}{2}\nabla_\mu X\right)\right)\,,\label{eq:PPD}
\end{align}
and
\begin{align}\label{eq:PNND}
{\cal P}^{\alpha\mu}{\cal N}^{\beta}{\cal N}^\nu{\cal D}_{\alpha\beta\mu\nu}&=\left({\cal P}^{\mu\nu}{\cal B}_{\mu\nu}\right)^2-{\cal P}^{\alpha\mu}{\cal P}^{\beta\nu}{\cal B}_{\alpha\beta}{\cal B}_{\mu\nu}
-\frac{2}{3}\frac{1}{\sqrt{X}}\nabla^\mu \left(\frac{1}{X^{3/2}}\left(\Box\phi\nabla_\mu\phi+\frac{1}{2}\nabla_\mu X\right)\right)\nonumber\\&-\frac{1}{3}\frac{1}{\sqrt{X}}\nabla^\mu\nabla^\nu \left(\frac{1}{X^{3/2}}\nabla_\mu\phi\nabla_\nu\phi\right)-\frac{1}{3}\frac{1}{\sqrt{X}}\nabla_\mu\nabla^\mu \left(\frac{1}{\sqrt{X}}\right)\,.
\end{align}
Note that in equations \eqref{eq:PPD} and \eqref{eq:PNND} we can do integration by parts in the action and the total derivatives can be written as being proportional to ${\cal P}^{\mu\nu}{\cal B}_{\mu\nu}$.
In principle it is possible to keep contracting such tensors indefinitely, e.g. $({\cal P}^{\mu\nu}\,{\cal B}_{\mu\nu})^n$, and the resulting quantity remains trivially disformal invariant.\\

\section{(U)-DHOST}
\label{sec:U-DHOST}
In this appendix we compare our formulation to the general U-DHOST. Let us consider a scalar-tensor theory up to quadratic order in second derivatives
\begin{align}
    S = \int \md^4x\,\sqrt{-g} \Big[ P + Q \Box \phi + f \mathcal{R} + C^{\mu\nu,\rho\sigma} \phi_{\mu\nu} \phi_{\rho\sigma} \Big]
\end{align}
where 
\begin{align}
    C^{\mu\nu,\rho\sigma} = \alpha_1 g^{\mu(\rho} g^{\sigma)\nu} + \alpha_2 g^{\mu\nu} g^{\rho\sigma} + \frac{1}{2} \alpha_3  \left( \phi^\mu\phi^\nu g^{\rho\sigma} + \phi^\rho \phi^\sigma g^{\mu\nu} \right)+ \alpha_4 \phi^{(\mu} g^{\nu)(\rho} \phi^{\sigma)} + \alpha_5 \phi^\mu\phi^\nu\phi^\rho \phi^\sigma
\end{align}
and $\alpha_i$, $P$, $Q$ and $f$ are function of $\phi$ and $X$. In general, these theories have four degrees of freedom (2 tensor and 2 scalar) due to the higher derivative terms. However, as discussed in \cite{Langlois:2015cwa,BenAchour:2016cay}, the additional degree of freedom can be eliminated by requiring degeneracy conditions between the free functions. The three conditions are given by
\begin{align}
    D_0(X)=& - 4 (\alpha_2 - \alpha_1) \Big[-X f (2\alpha_1 - X \alpha_4 - 4 f_X) - 2 f^2 - 8 X^2 f_X^2 \Big], \\
    D_1(X) =& 4 \alpha_4 \left( X^2 \alpha_1 (\alpha_1 + 3 \alpha_2) - 2 f^2 + 4 X f \alpha_2 \right)  + 4 X^2 f (\alpha_1 + \alpha_2) \alpha_5 - 8 X \alpha_1^3 \nonumber \\
    &- 4 \alpha_1^2 ( f + 4 X f_X + 6 X \alpha_2 ) \alpha_1^2- 16 (f + 5 X f_X) \alpha_1 \alpha_2 - 4 X (3f - 4 X f_X)\alpha_1 \alpha_3 \nonumber \\
    &- X^2 f \alpha_3 - 32 f_X (f + 2 X f_X) \alpha_2 + 16 f f_X \alpha_1 - 8 f ( f- X f_X) \alpha_3 + 48 f f_X^2, \\
    D_2(X)=& 4 \left( 2 f^2 - 4 X f \alpha_2 - X^2 \alpha_1 (\alpha_1 + 3\alpha_2) \right) \alpha_5  + 4\alpha_1^3 + 4 (2\alpha_2 + X \alpha_3 + 4 f_X) \alpha_1^2 + 3 X^2 \alpha_1 \alpha_3^2 \nonumber \\
    &+ 4 X f \alpha_3^2 + 8 (f + X f_X) \alpha_1 \alpha_3 + 32 f_X \alpha_1 \alpha_2 + 16 f_X^2 (\alpha_1 + 2\alpha_2) + 16 f f_X \alpha_3\,.
\end{align}
If $D_0=D_1=D_2=0$ the higher time derivative terms are canceled and the theory contains just three degrees of freedom. Mimetic gravity can be understood as a subclass of DHOST since it only has three degrees of freedom despite the presence of higher derivative terms and it indeed fulfills the above degeneracy conditions \cite{Langlois:2018jdg}.

The previous degeneracy conditions apply for any form of the scalar field $\phi$. However, if the scalar field has a uniform slicing $\phi=\phi(t)$ terms proportional to the scalar gradient $D_i \phi$ vanish identically. Therefore, in the unitary gauge the DHOST class can be enlarged to the U-DHOST class \cite{DeFelice:2018ewo} and the degeneracy conditions are now given by
\begin{align}
    \alpha_1 =& \kappa_1 - \frac{f}{X}, \qquad \alpha_2 = \kappa_2 + \frac{f}{X}, \qquad \alpha_3 = 2 \frac{f}{X^2} - 4 \frac{f_X}{X} + 2 \sigma \kappa_1 + 2 \left( 3\sigma + \frac{1}{X} \right)\kappa_2, \nonumber \\
    \alpha_4=& \alpha + 2 \frac{f_X}{X} - 2 \frac{f}{X^2} + \frac{2}{X} \kappa_1, \qquad \alpha_5=\frac{\alpha}{X}- 2 \frac{f_X}{X^2}+ \kappa_1 \left(\frac{1}{X}+3 \sigma^2+ \frac{2\sigma}{X} \right) + \kappa_2 \left( 3\sigma + \frac{1}{X}\right)^2\,,
\end{align}
where $\sigma$, $\kappa_i$, and $\alpha$ are functions of $X$ and $\phi$.

One can check that the special disformal symmetric model \eqref{eq:actionweuse} corresponds to
\begin{align}
P&=\sqrt{X}\left(c_0+c_{1,\phi}+2Xc_{2,\phi\phi}\right)\,,\\
Q&=-4c_{2,\phi}\sqrt{X}\,,\\
f&=c_2\sqrt{X}\\
\alpha_1&=\sqrt{X}\left(-\frac{c_2}{X}+\frac{c_4}{X^2}\right)\,,\\
\alpha_2&=\sqrt{X}\left(\frac{c_2}{X}+\frac{c_3}{X^2}\right)\,,\\
\alpha_3&=2\sqrt{X}\frac{c_3}{X^3}\,,\\
\alpha_4&=\sqrt{X}\left(2\frac{c_4}{X^3}+\frac{d_1}{X^2}\right)\,,\\
\alpha_5&=\sqrt{X}\left(\frac{d_1}{X^3}+\frac{c_4}{X^4}+\frac{f_1}{X^4}\right)\,.
\end{align}
In the case where $d_1=f_1=0$ then we go to the generic disformal symmetric action \eqref{eq:actioncovariant}.

\bibliography{bibliography}

\end{document}